\documentclass[aps,prl,reprint,superscriptaddress,twocolumn]{revtex4-1}
\usepackage{graphicx}% Include figure files
\usepackage[english]{babel} % 
\usepackage[usenames]{color}
\usepackage{subfigure}
\usepackage{dcolumn}% Align table columns on decimal point
\usepackage{bm}% bold math
\usepackage{amsmath,amsfonts,amssymb,float}
\usepackage{natbib}
\usepackage{epsfig}
\usepackage{multirow}
\usepackage{textcomp} % allows upright mu s%
\usepackage{amsmath} % AMS Math Package
\usepackage{amsthm} % Theorem Formatting
\usepackage{amssymb}	% Math symbols such as \mathbb
\usepackage{graphicx} % Allows for eps images
\makeatletter % Need for anything that contains an @ command 
\let\baraccent=\= % rename builtin command \= to \baraccent
\renewcommand{\=}[1]{\stackrel{#1}{=}} % for putting numbers above =

\theoremstyle{definition}

\theoremstyle{remark}

\begin{document}

% Use the \preprint command to place your local institutional report
% number in the upper righthand corner of the title page in preprint mode.
% Multiple \preprint commands are allowed.
% Use the 'preprintnumbers' class option to override journal defaults
% to display numbers if necessary
%\preprint{}

%Title of paper
\title{Numerical Modeling of the Sensitivity of X-Ray Driven Implosions to Low-Mode Flux Asymmetries}

\author{R.H.H. Scott}
\email{Robbie.Scott@stfc.ac.uk}
\thanks{Also visiting scientist at Lawrence Livermore National Laboratory, Livermore, CA 94551, United States of America, \& Department of Physics, The Blackett Laboratory, Imperial College London, Prince Consort Road, London, SW7 2AZ, United Kingdom. The authors thank M.H. Key and C. Cerjan for useful discussions, the staff of NIF and Livermore Computing.}

\affiliation{Central Laser Facility, STFC Rutherford Appleton Laboratory, Harwell Oxford, Didcot, OX11 0QX, United Kingdom}
\author{D.S. Clark}
\author{D.K. Bradley}
\author{D.A. Callahan}
\author{M.J. Edwards}
\author{S.W. Haan}
\author{O.S. Jones}
\author{B.K. Spears}
\author{M.M. Marinak}
\author{R.P.J. Town}
\affiliation{Lawrence Livermore National Laboratory, Livermore, CA 94551, United States of America}
\author{P.A. Norreys}
\affiliation{Central Laser Facility, STFC Rutherford Appleton Laboratory, Harwell Oxford, Didcot, OX11 0QX, United Kingdom}
\affiliation{Department of Physics, University of Oxford, Parks Road, Oxford OX1 3PU, United Kingdom}
\author{L.J. Suter}
\affiliation{Lawrence Livermore National Laboratory, Livermore, CA 94551, United States of America}
\date{\today}

\begin{abstract}
The sensitivity of inertial confinement fusion implosions of the type performed on the National Ignition Facility (NIF) \cite{lindl:339} to low-mode flux asymmetries has been investigated numerically. It is shown that large-amplitude, low-order mode shapes (Legendre polynomial $P_4$), resulting from associated low order flux asymmetries, cause spatial variations in capsule \& fuel momentum that prevent the DT ``ice'' layer from being decelerated uniformly by the hot spot pressure. This reduces the transfer of kinetic to internal energy of the central hot spot, thus reducing neutron yield. Furthermore, synthetic gated x-ray images of the hot spot self-emission indicate that $P_4$ shapes may be unquantifiable for DT layered capsules. Instead the positive $P_4$ asymmetry ``aliases'' itself as an oblate $P_2$ in the x-ray self emission images. Correction of this apparent $P_2$ distortion can further distort the implosion while creating a round x-ray image. Long wavelength asymmetries may be playing a significant role in the observed yield reduction of NIF DT implosions relative to detailed post-shot 2D simulations.

%Based on $P_4$ mode amplitudes measured on recent NIF shots, this modeling suggests that equivalent spherical implosions may result in significantly enhanced neutron yields.
\end{abstract}
% insert suggested PACS numbers in braces on next line
% insert suggested keywords - APS authors don't need to do this
%\keywords{}
%\maketitle must follow title, authors, abstract, \pacs, and \keywords
\maketitle

Indirect-drive inertial confinement fusion (ICF) \cite{Basov:1991uq,NUCKOLLS:1972fk,lindl:339} uses lasers to heat the inside of a cavity (or hohlraum). The absorbed laser energy is re-emitted as approximately black-body radiation in the soft x-ray regime. These x-rays heat the outer surface of a hollow, spherical, $\sim 2$ mm diameter, shell that contains a $\sim 70$ \textmu m thick layer of cryogenically frozen Deuterium and Tritium fuel (``DT fuel'' or ``DT layered capsules''). The heated outer shell ablates, which creates a reaction force, accelerating the remaining shell spherically inwards at extremely high velocity ($\sim 350$ km/s). During the implosion, spherical convergence causes the pressure in the gaseous void (or hot spot) within the shell to rise rapidly. This pressure decelerates the shell, simultaneously compressing the solid fuel and converting the shell's kinetic energy into hot spot internal energy. If this conversion rate exceeds loss rates due to thermal conduction and bremsstrahlung radiation, the hot spot will heat, initiating DT fusion reactions. Provided the hot spot areal density is sufficient, $\alpha$-particles created by the fusion reactions will redeposit their energy locally, further heating the hot spot, resulting in bootstrap heating, ignition, and propagation of burn into the surrounding cold fuel. Numerical modeling indicates that the National Ignition Facility (NIF) can, for the first time, initiate inertial fusion ignition in the laboratory \cite{haan:051001,clark:052703,1742-6596-112-2-022021}.

In this Letter, the effects of large, low-mode asymmetries in the x-ray drive are examined numerically. The non-uniformity of the x-ray flux incident upon the shell and the resultant non-spherical shell shapes can be described mathematically as a series of Legendre polynomials \cite{Abramowitz:1972kx}. It is shown that a large-amplitude $P_4$ implosion asymmetry, that might result from low-order hohlraum generated flux asymmetries, causes spatial variations in the capsule \& fuel momentum. This can inhibit the DT fuel from being decelerated uniformly by the hot spot pressure, reducing the efficient transfer of implosion kinetic energy to hot spot internal energy thus significantly reducing the capsule performance. Furthermore, simulated gated x-ray images of the hot spot self-emission show reduced sensitivity to the $P_4$ mode, instead the images appear to have a pronounced oblate $P_2$ shape. Reducing the amplitude of the oblate $P_2$ shape (as measured from the x-ray image) further reduces the sensitivity to the $P_4$ mode such that no quantitative evaluation of the hot spot $a_4$ (where $a_4$ is the amplitude of the $P_4$ mode) can be made, furthermore the x-ray images are circular despite the capsule shape being highly distorted. Comparisons are made between key physical properties of the implosion, synthetically generated experimental observables, and NIF experimental data.

\begin{figure}[t!]
\centering
\subfigure[]{}\includegraphics[scale=.19,trim=20mm 105mm 40mm 40mm,clip,angle=0]{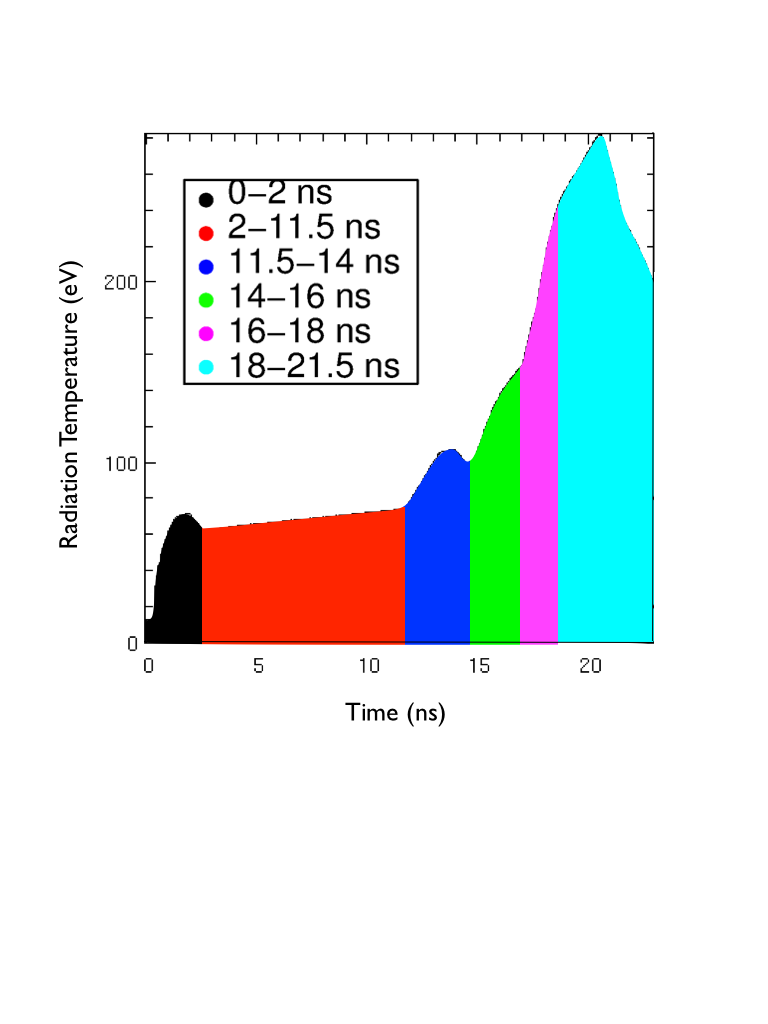}
\subfigure[]{}\includegraphics[scale=.28,trim=5mm 5mm 30mm 20mm,clip]{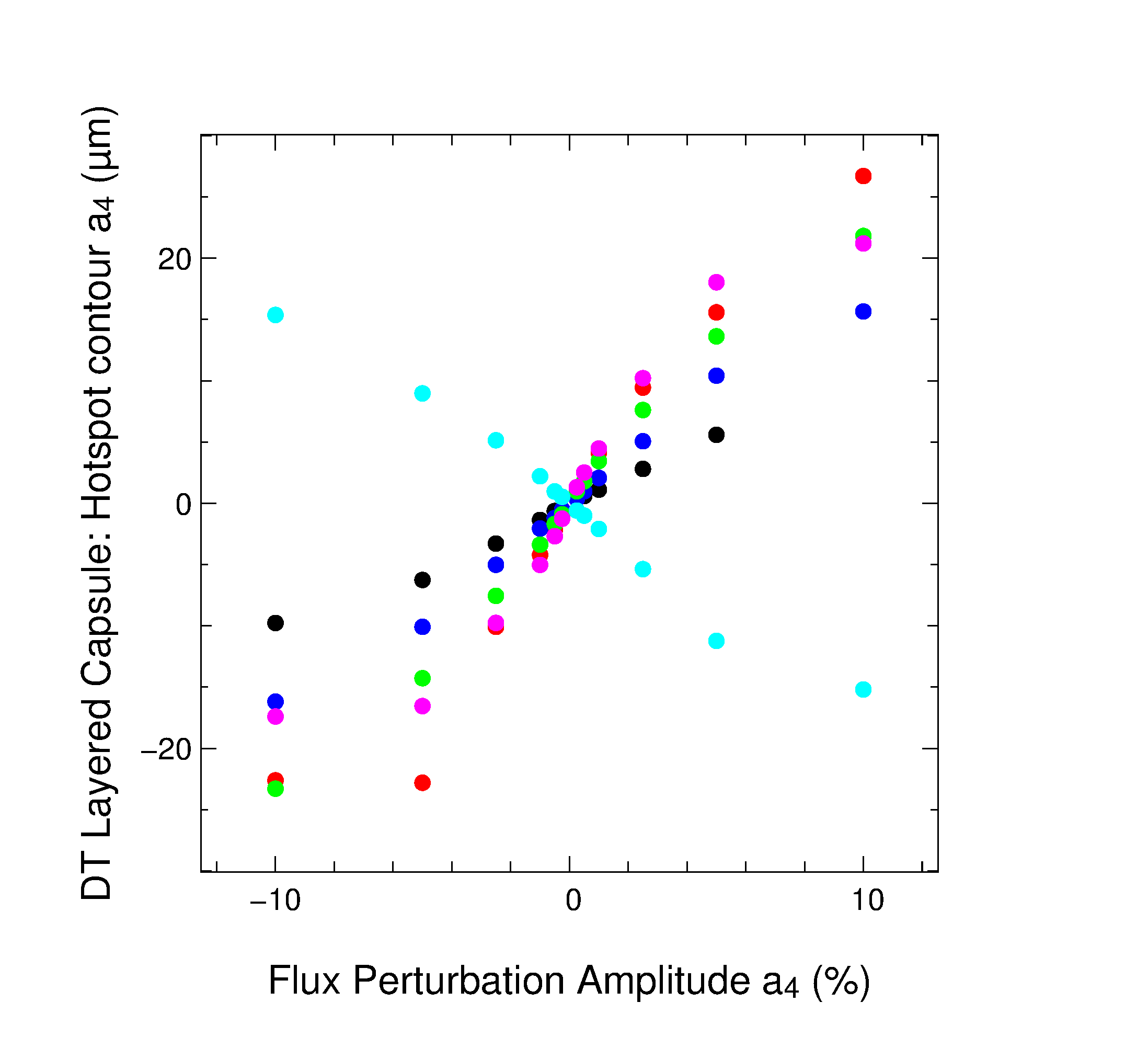}
\subfigure[]{}\includegraphics[scale=.28,trim=5mm 5mm 30mm 20mm,clip]{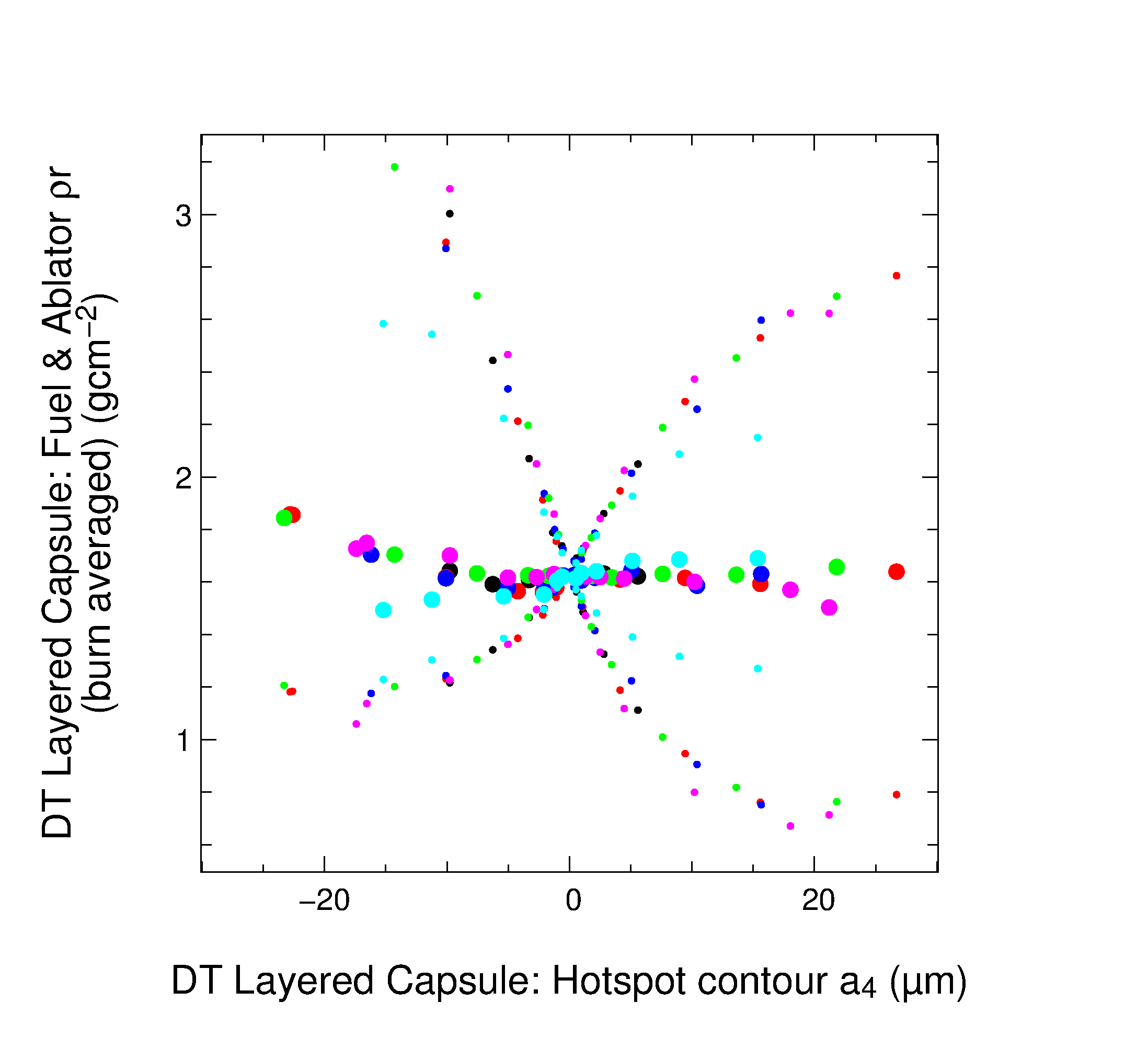}
\subfigure[]{}\includegraphics[scale=.28,trim=5mm 5mm 30mm 20mm,clip]{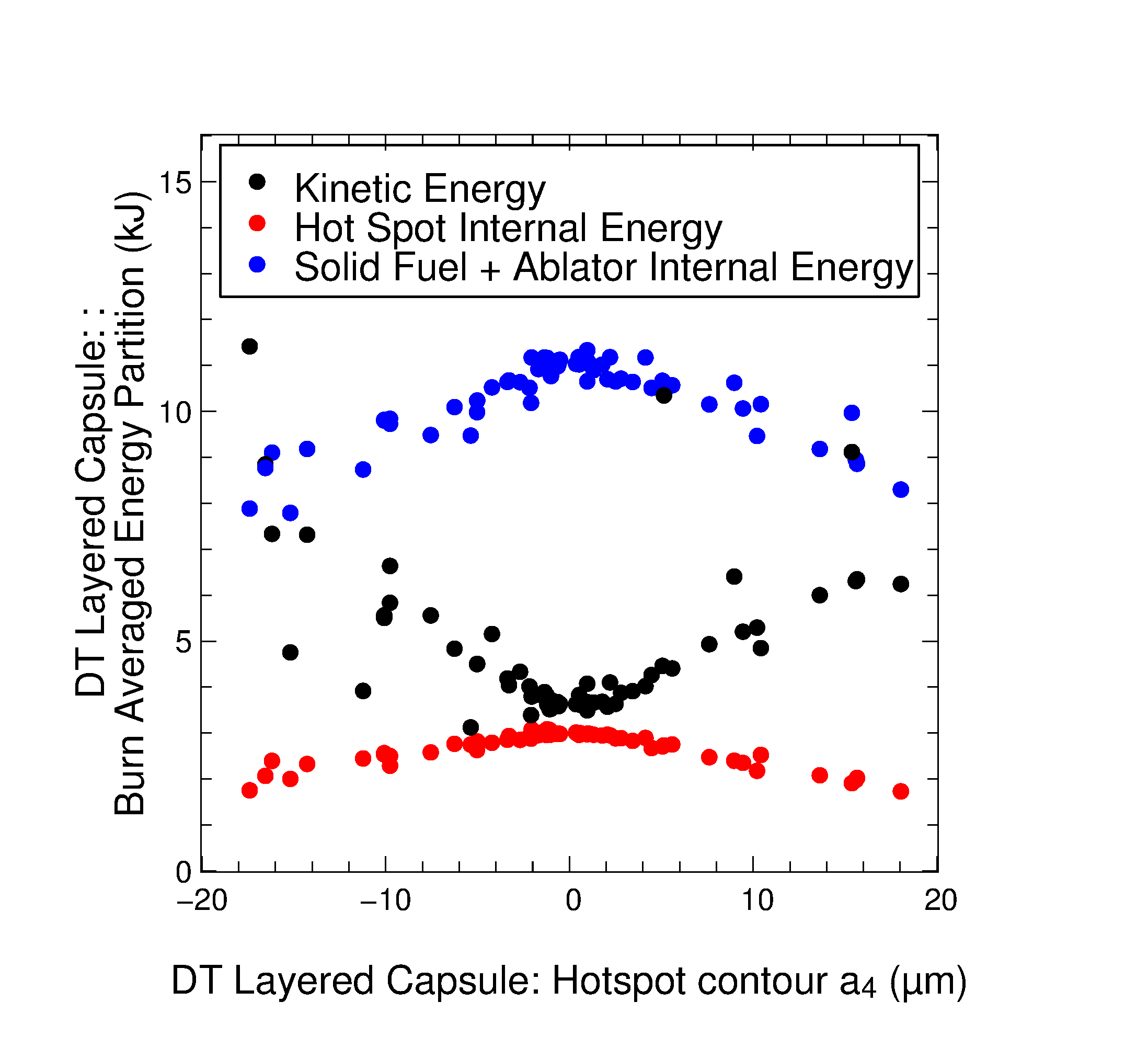}
\subfigure[]{}\includegraphics[scale=.28,trim=5mm 5mm 30mm 20mm,clip]{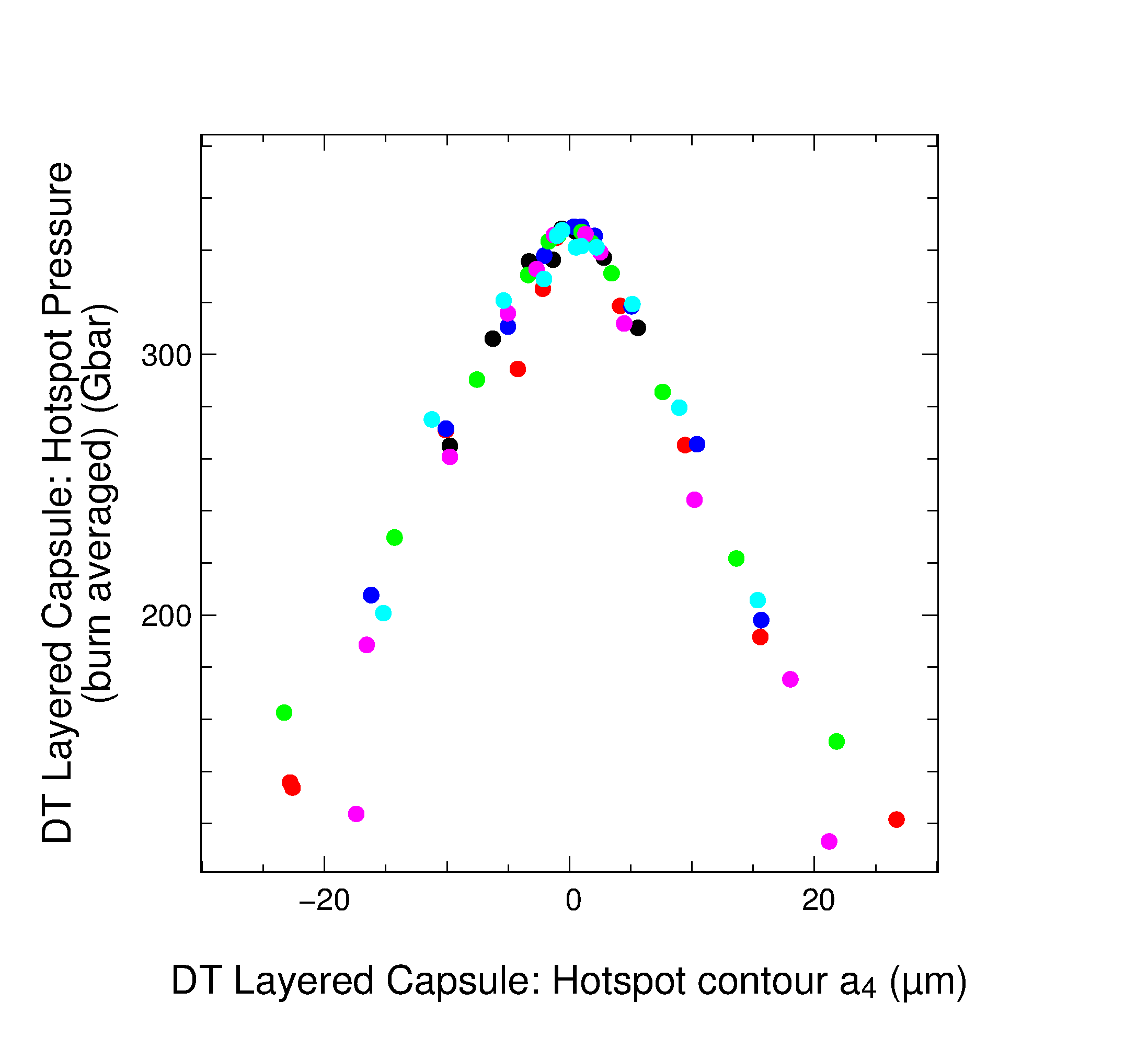}
\subfigure[]{}\includegraphics[scale=.28,trim=5mm 5mm 30mm 20mm,clip]{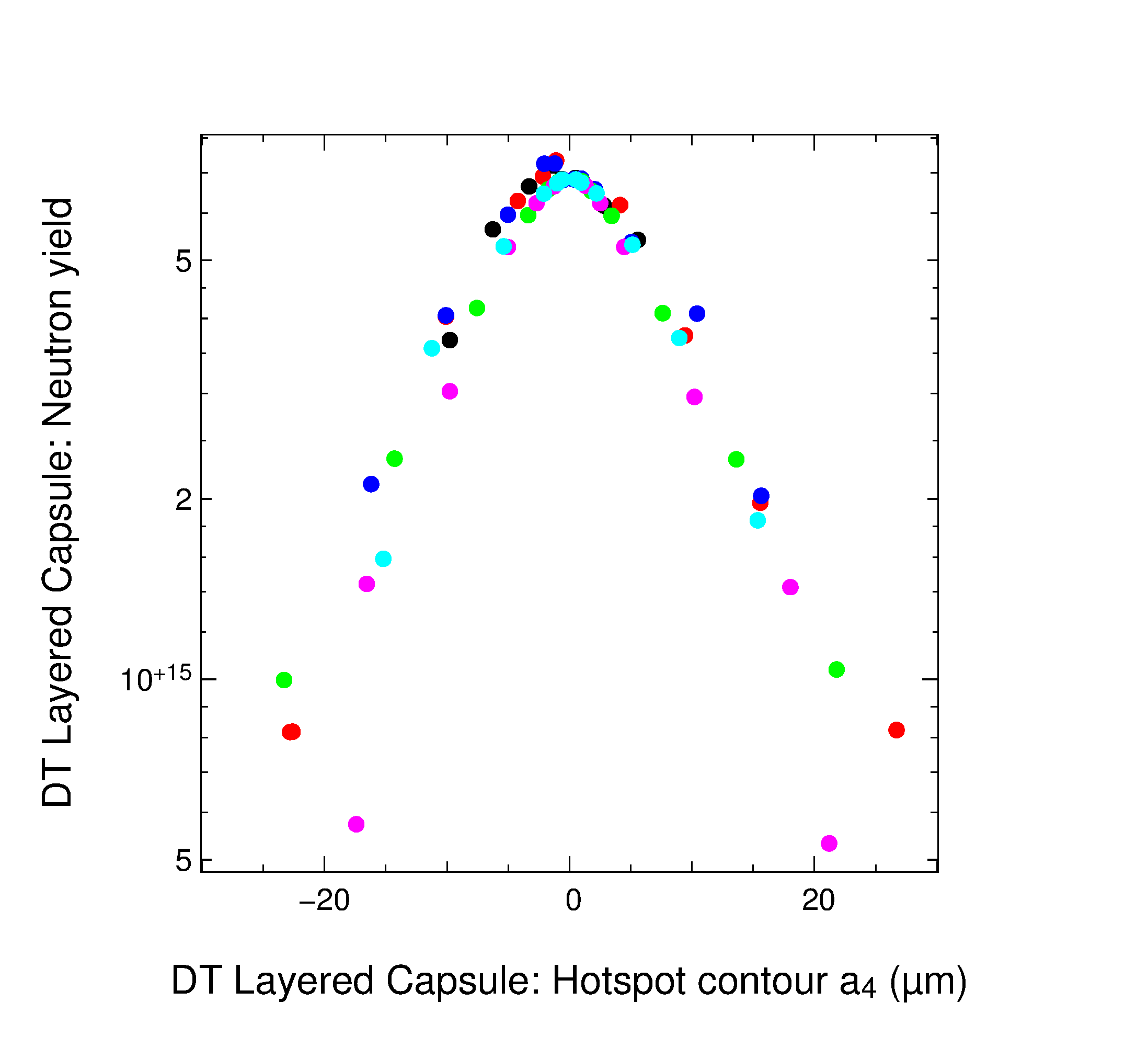}
%\subfigure[]{}\includegraphics[scale=.28,trim=0mm 0mm 0mm 0mm,clip]{./DTmassHS_DTp4myC.png}
%\subfigure[]{}\includegraphics[scale=.28,trim=0mm 0mm 0mm 0mm,clip]{./DTvolumeHS_DTp4myC.png}
	\caption{(a) Applied radiation temperature vs time. Colors depict when during the drive pulse the various flux asymmetries were applied as shown on subsequent plots. (b) The relationship between applied flux asymmetry and hot spot $a_4$ at x-ray bangtime (the time at which peak emission occurs). Each point represents a 2D Hydra run. (c) Burn averaged hot spot + fuel + ablator $\rho r$ vs hot spot $a_4$ at x-ray bangtime: large dots are spatially averaged $\rho r$, while the smaller points with the same color and $a_4$ are the maxa and minima of the spatially averaged value; large variations in $\rho r$ occur due to $P_4$. (d)  The burn averaged energy partition as a function of hot spot $a_4$; increasing $P_4$ perturbations prevent kinetic energy (black) from being converted to both hot spot (red) and solid fuel (blue) internal energy during stagnation. (e) Burn averaged hot spot pressure as a function of hot spot $a_4$. (f) Total thermonuclear neutron yield as a function of hot spot $a_4$; yield varies by a factor of 15. }%(e) hot spot mass as a function of hot spot $a_4$. (f) hot spot volume as a function of hot spot $a_4$.
\label{fig2}
\end{figure}

The indirect-drive approach to ICF smooths high mode spatial non-uniformities in the x-ray flux incident on the capsule, however the spatial distribution of the cones of laser beams which illuminate the hohlraum means that low mode x-ray flux non-uniformities can occur \cite{lindl:339}, these are considerably lower mode than those recently examined by Thomas \emph{et al} \cite{PhysRevLett.109.075004}. The growth of Legendre polynomial $P_4$ capsule shapes was investigated using the radiation-magnetohydrodynamics code Hydra \cite{marinak:2275}. A frequency and time dependent x-ray source was developed to drive these capsule-only simulations. The initial x-ray drive was taken from an integrated hohlraum simulation and then adjusted to match the shock timing data obtained using the VISAR diagnostic \cite{barker:4669} from NIF shot N110521, and the capsule implosion trajectory \cite{hicks:102703} measured on NIF shot N110625. A $90^{\circ}$ `wedge' of the capsule ($2\pi$ Sr) was modeled using two-dimensional (2D) cylindrically-symmetric geometry with $256\times312$ cells. Doubling and quadrupling the cell resolution demonstrated convergence. In all runs the Quotidian Equation of State \cite{more:3059} was used with tabular opacities and multi-group radiation diffusion. The effects of Legendre polynomial $P_4$ hohlraum flux asymmetries were investigated by perturbing the tuned x-ray drive with spatially varying flux asymmetries of the form: $fds(\theta,t)=(a_0P_0 + a_4P_4(\theta))*fds(t)$ where $fds$ is the energy density of the tuned photon frequency dependent x-ray drive source, $a_n$ is the amplitude of the $n^{th}$ Legendre polynomial, $a_0=1$, $a_4= (\pm0.10, \pm0.05, \pm0.025, \pm0.01, \pm0.005, \pm0.0025)$, $\theta$ the angle between the equatorial plane and polar axis, and $t$ time. Hydra modeling of the hohlraum \& capsule for nominal implosions suggests the flux asymmetry incident on the capsule would be expected to vary by $<3$\% except for in the first $\sim 2$ ns of the laser pulse were it can vary by up to 10\% \cite{Jones:2012}. The flux asymmetries were applied 100 \textmu m from the capsule ablation front during discrete time intervals (see fig. \ref{fig2}(a)), creating a database of $>200$ 2D modeling runs of both DT layered implosions and DHe$^3$ gas filled capsules with a surrogate inner CH layer of equal mass to a DT fuel layer (symmetry capsules). Time resolved synthetic gated x-ray images of the hot spot self-emission $>6$ keV, including its attenuation by the compressed fuel and ablator, were created from both polar and equatorial directions by post processing each Hydra run. The images were blurred in order to reproduce the $11$ \textmu m resolution of the diagnostic. The key implosion performance metrics (neutron yield, hot spot pressure, mass, volume, density, ion \& electron temperatures, the effective ion temperature computed from the FWHM of the DT neutron spectrum, fuel and ablator areal density ($\rho r$) and kinetic energy) were extracted from the simulations. The hot spot shape was evaluated as a function of time by performing a Legendre polynomial decomposition (modes 1-10) of the appropriate contour. For DT layered capsules the hot spot contour is defined for each angular `strip' of cells $j$ as the minimum radius where $T_{e_j} > \frac{1}{2}T_{e_{j_{max}}}$ and $\rho_j < \frac{1}{2}\rho_{j_{max}}$ where $T_{e}$ is the electron temperature and $\rho$ the mass density, `max' denotes the maximum value within the $j^{th}$ strip. This has been found to produce a robust definition of the hot spot even for highly distorted implosions.  The 17\% contour of the gated x-ray diagnostic (GXD) is used both for the synthetic GXD and experimentally, as previous studies have shown this provides a faithful representation of the hot spot shape for small departures from sphericity.

\begin{figure}[t!]
\centering
\subfigure[]{}\includegraphics[scale=.14,trim=0mm 10mm 48mm 0mm,clip,angle=0]{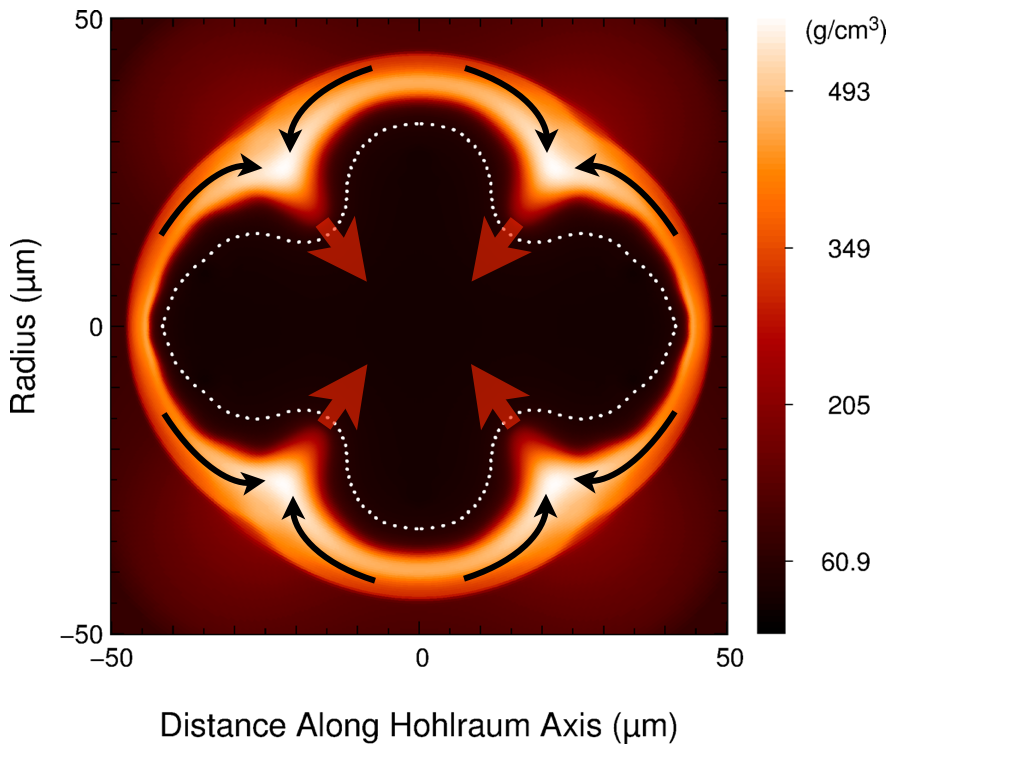} 
\subfigure[]{}\includegraphics[scale=.25,trim=0mm 0mm 0mm 0mm,clip,angle=0]{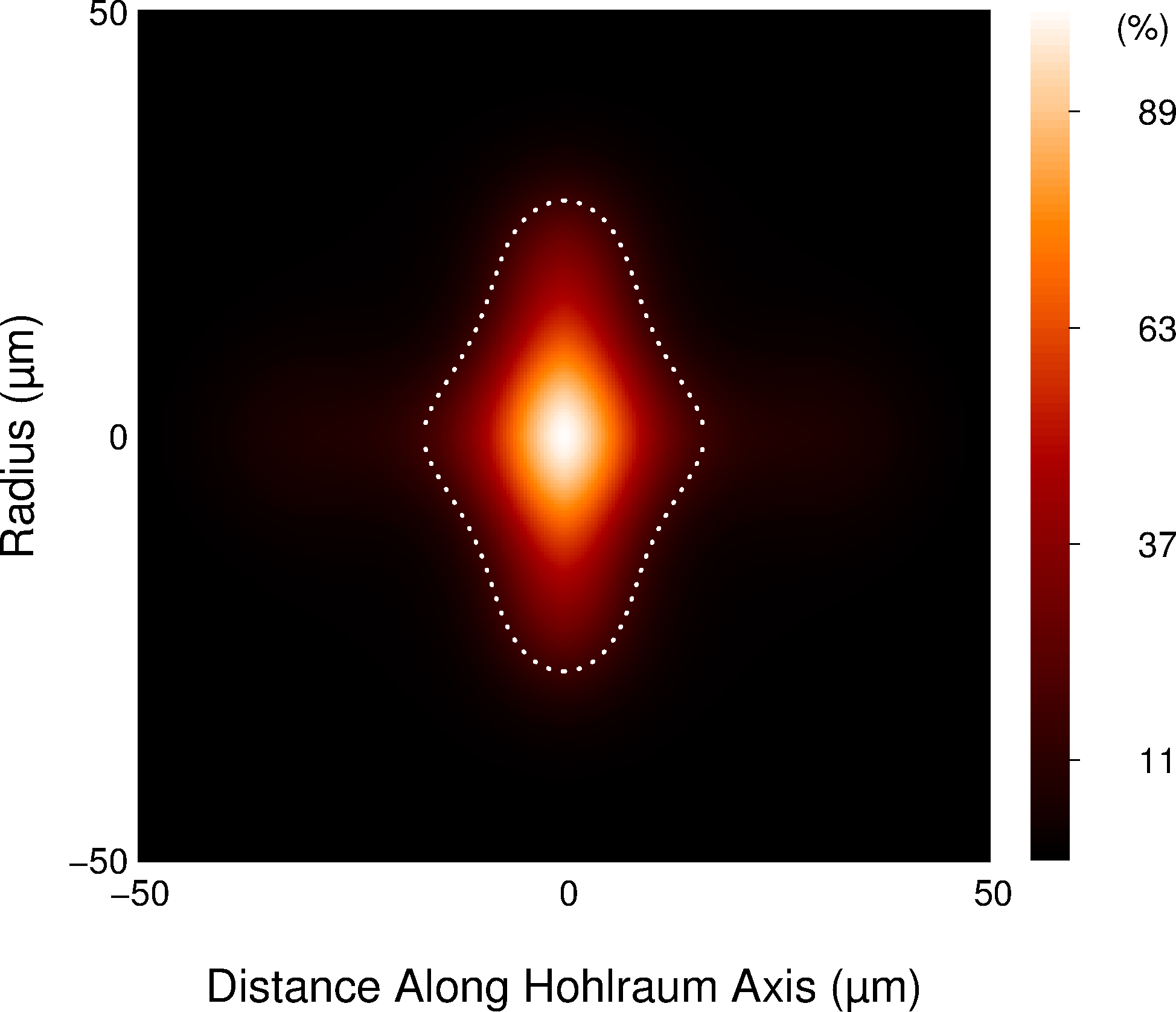}
\subfigure[]{}\includegraphics[scale=.25,trim=0mm 0mm 0mm 0mm,clip]{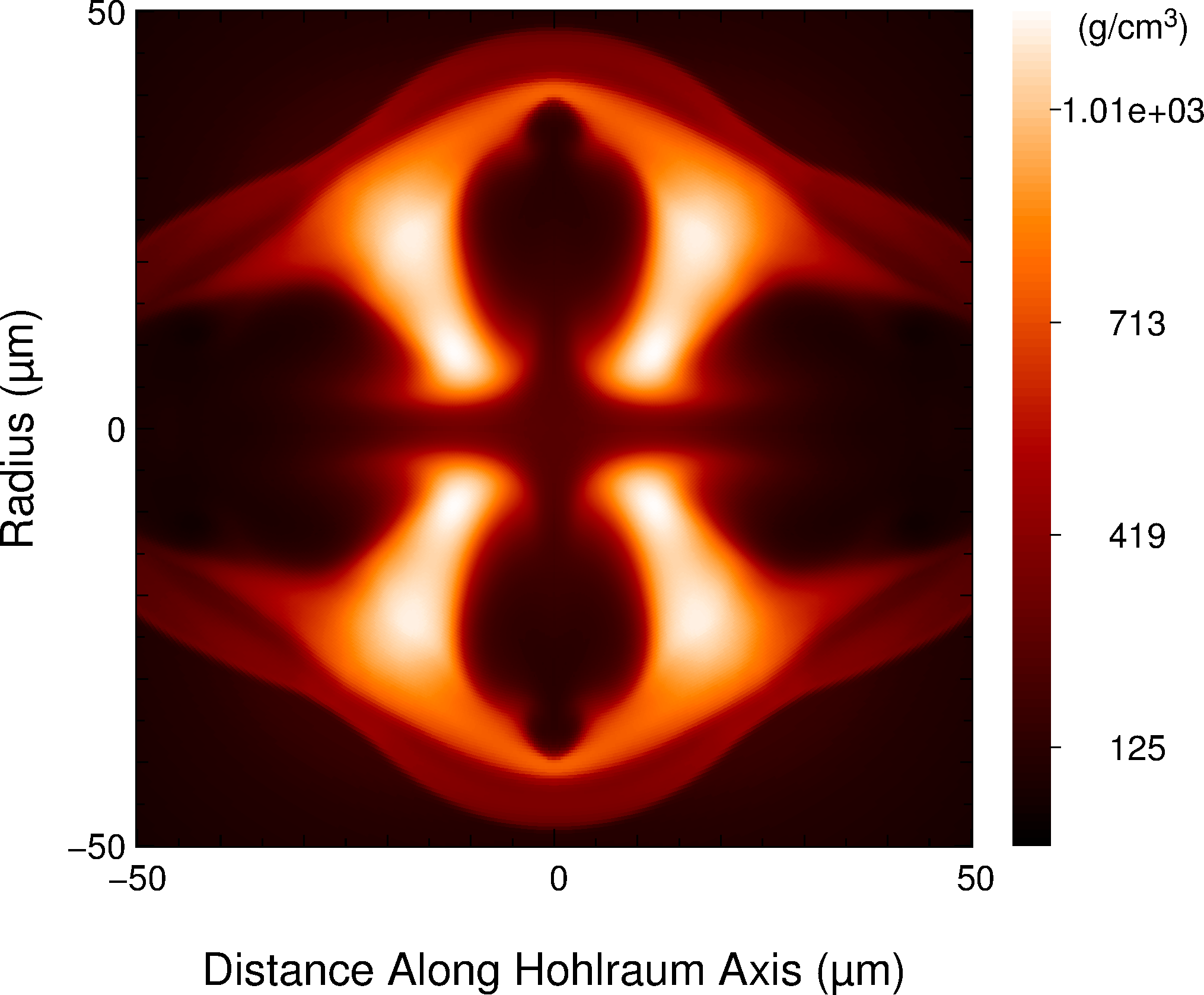}
\subfigure[]{}\includegraphics[scale=.25,trim=0mm 0mm 0mm 0mm,clip]{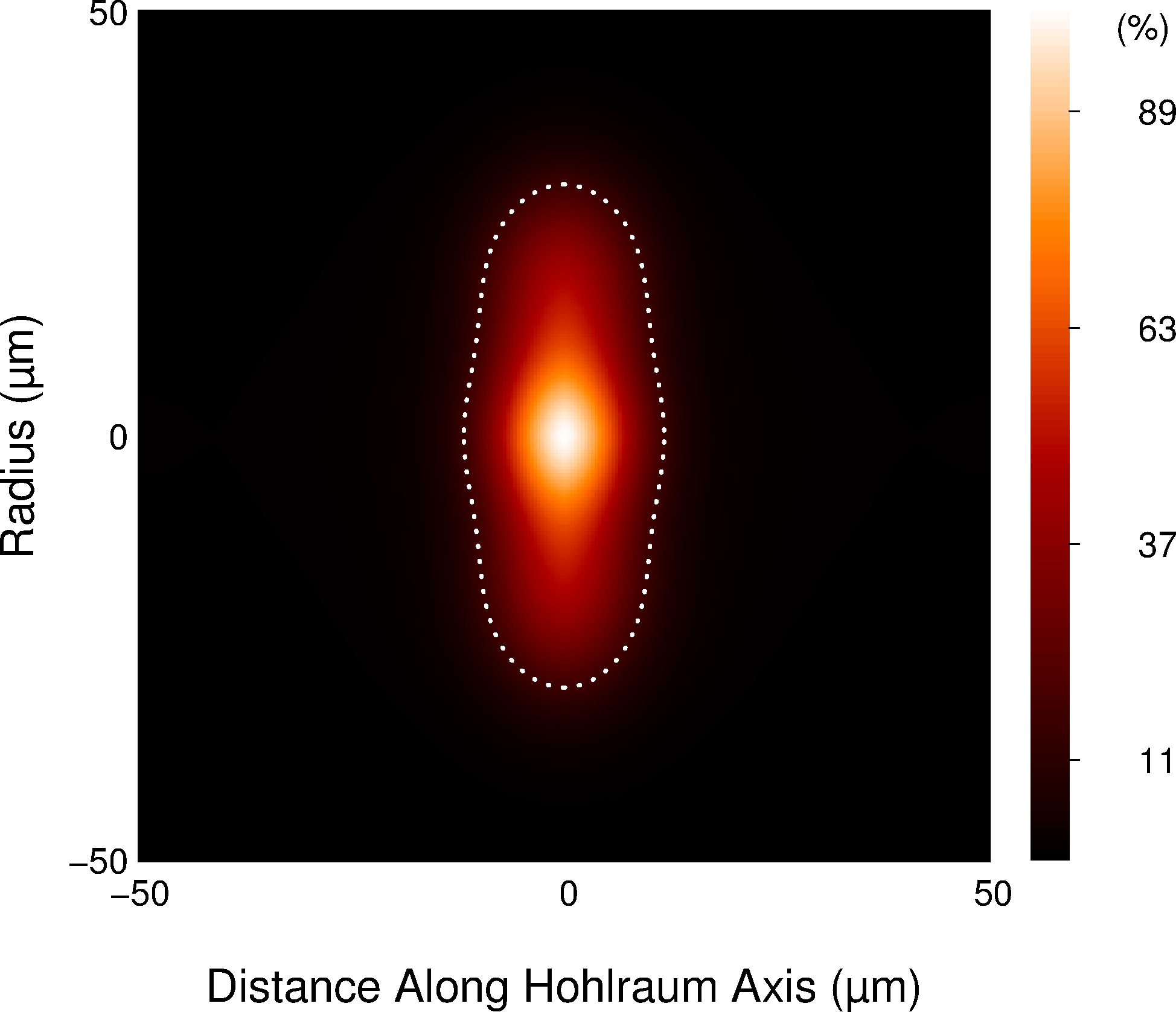}
\caption{Axis of rotational symmetry is horizontal at Radius $=0$ \textmu m. (a) DT layered capsule density plot at x-ray bangtime showing a positive Legendre polynomial $P_4$ shape. This simulation had a 10\% flux asymmetry applied from 11.5-14 ns. Black arrows indicate the mass flows which occur during stagnation. After bangtime `fingers' of fuel continue to flow inwards (red arrows). White dots depict the hot spot contour. (b) Synthetic gated x-ray image of the hot spot self emission from (a), white dots show the 17\% contour. (c) Fig. (a) 100 ps later: due to burn truncation with large $a_4$ this is the neutron bangtime for an equivalent spherical implosion. (d) The synthetic GXD from (c), showing a large negative $P_2$ and almost zero $a_4$ despite the obvious $P_4$ in (c).}
\label{picture}
\end{figure}

The applied Legendre $P_4$ flux asymmetries induce $P_4$ hot spot shapes at stagnation (see Figs. \ref{picture} (a) \& (c)), the sign of which is dependent on the timing of the applied flux asymmetry. If the asymmetry is present only during the shock compression phase (the first $\sim 18$ ns), shocks created in regions of the capsule exposed to higher flux propagate faster, these faster shocks break out of the inner DT ice layer earlier, causing these regions to move ahead of those exposed to less flux. This also causes ablator mass to flow laterally, away from the high flux region. Consequently during peak drive the regions initially exposed to high flux are at smaller radii, meaning they are accelerated less efficiently by the hohlraum flux and gain less total momentum. They can also have less ablator aerial density. The net effect is that the regions experiencing high flux during shock compression will protrude outwards at stagnation. Conversely if the flux asymmetry is applied during peak drive, the regions of the capsule exposed to more flux gain more momentum, and protrude inwards at stagnation. Regardless of the timing of the applied asymmetry, during the stagnation phase of the implosion, pressure within the lower density hot spot decelerates the higher density fuel from peak velocity, making this interface Rayleigh-Taylor unstable \cite{Rayleigh:1900kx,Taylor:1950vn}. The instability will accentuate any shape imperfections during deceleration, as indicated by the significant simulated growth shown in figs. \ref{picture}(a) \& (c). 

The scalings of some important DT layered capsule implosion parameters as a function of hot spot $a_4$ measured at x-ray bangtime are summarized in figure \ref{fig2}. Fig. \ref{fig2}(b) shows the relationship between applied $P_4$ flux perturbation amplitude and the resulting shape $a_4$ at x-ray bangtime. Fig. \ref{fig2}(c) depicts the `burn averaged' $\rho r$ (the burn average of a quantity $Q_b = (\sum_{t=0}^{t=\infty} Q_t Eprodr dt)/\int_{t=0}^{t=\infty}Eprodr\ dt$ where $Q_t$ is Q at time $t$ and $Eprodr$ the thermonuclear energy production rate in time $dt$) as a function of hot spot $a_4$. Although the spatially averaged $\rho r$ is relatively constant, the lateral mass flows caused by the $P_4$ can create large spatial variations in $\rho r$. The regions with higher momentum continue to propagate radially inwards; fig. \ref{fig2}(d) depicts the remaining capsule kinetic energy as a function of $a_4$ and the partition of that energy into hot spot and solid fuel internal energy. For large $a_4$ less of the implosion kinetic energy is converted into hot spot internal energy and the hot spot pressure is reduced (see fig. \ref{fig2}(e)). The reduction in neutron yield can be as large as 15$\times$ for hot spot $a_4=20$ \textmu m (flux asymmetry $\sim 10\%$) as shown in fig. \ref{fig2}(f)).  

%\begin{figure}[t!]
%\centering
%\subfigure[]{}\includegraphics[scale=.28,trim=0mm 0mm 0mm 0mm,clip]{./DTkeshell_DTyield.png}
%\subfigure[]{}\includegraphics[scale=.28,trim=0mm 0mm 0mm 0mm,clip]{./DTprhs_DTyield.png}
%\subfigure[]{}\includegraphics[scale=.28,trim=0mm 0mm 0mm 0mm,clip]{./DTnyield_DTp4_myC.png}
%	\caption{(a) Total thermonuclear neutron yield as a function of hot spot pressure averaged over the hot spot volume and thermonuclear burn duration. (b) Total thermonuclear neutron yield as a function of hot spot $a_4$.}
%\label{fig:yieldke_yieldpr}
%\end{figure}
%fig 3
\begin{figure}[t!]
\centering
\subfigure[]{}\includegraphics[scale=.28,trim=5mm 5mm 30mm 16mm,clip]{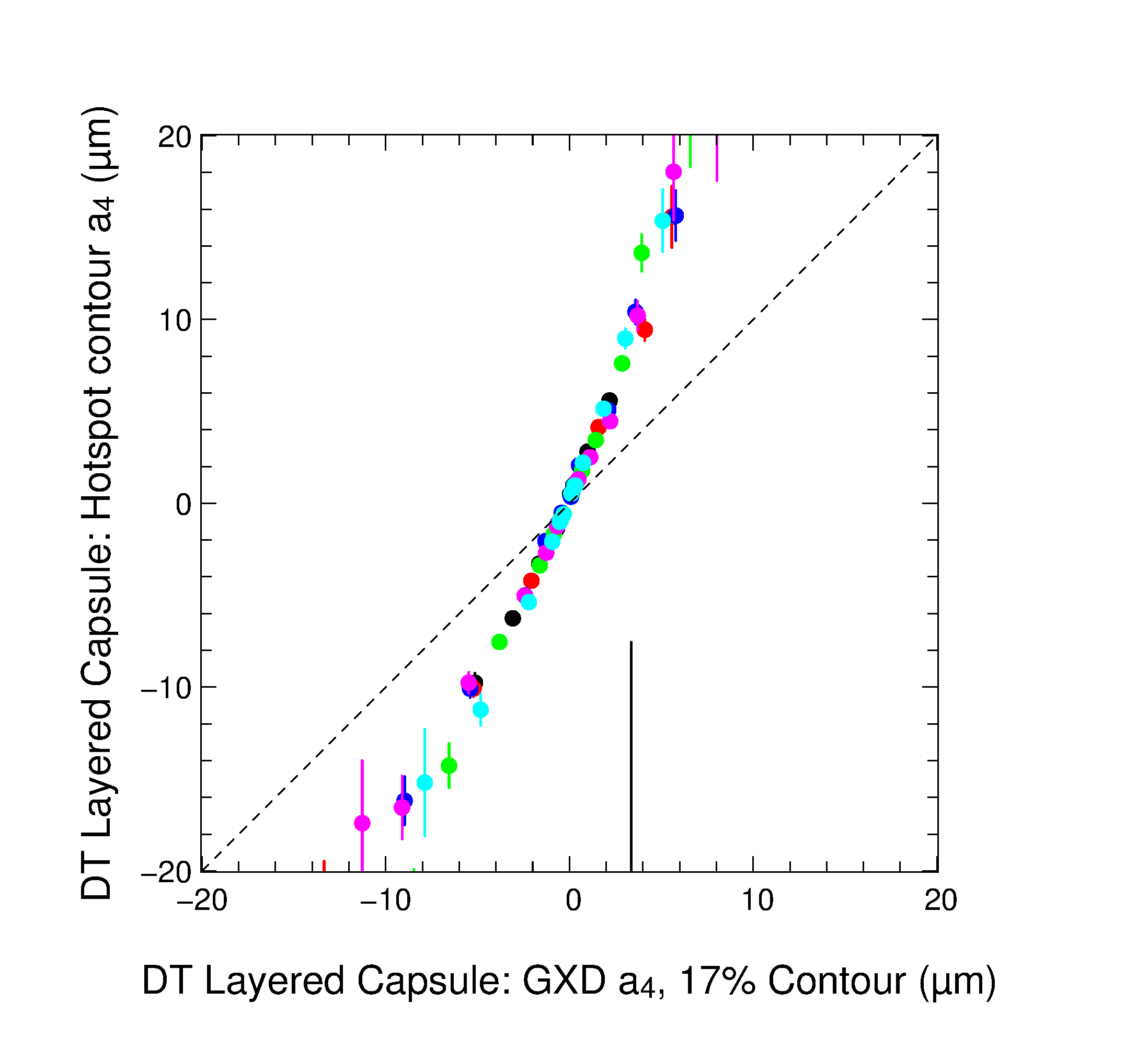}
\subfigure[]{}\includegraphics[scale=.28,trim=5mm 5mm 30mm 16mm,clip]{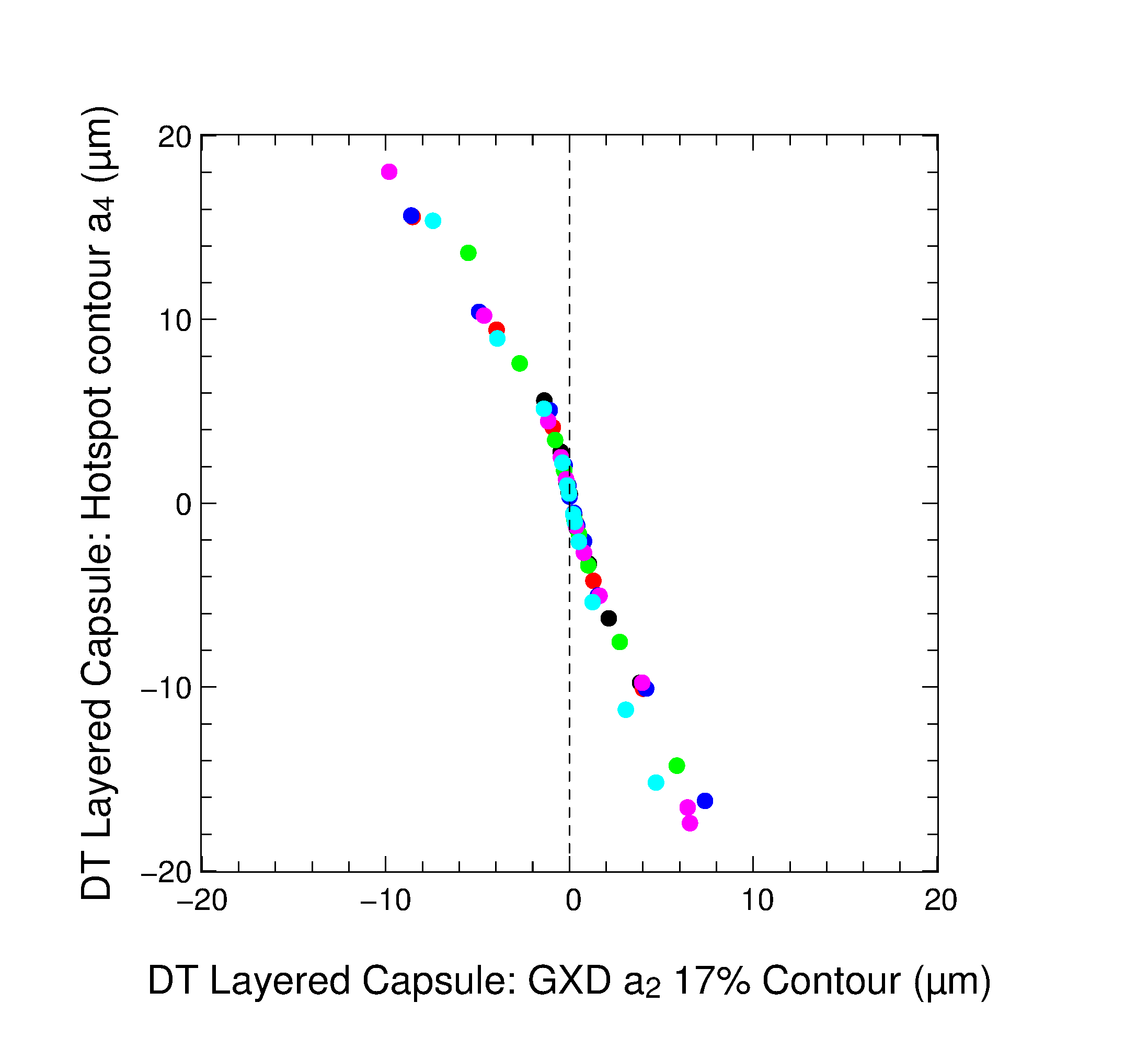}
\caption{(a) DT layered capsule hot spot $a_4$ plotted against the synthetic GXD $a_4$; particularly for large positive $a_4$ the GXD is unable to effectively measure the amplitude of the $P_4$ mode. (b) DT layered capsule hot spot $a_4$ plotted against the synthetic GXD $a_2$; the GXD measures a significant $P_2$ mode amplitude despite the DT layered capsule hot spot $a_2$ being $0\pm1$  \textmu m (not shown).}
\label{fig3}
\end{figure}
%This insensitivity is compounded when realistic experimental noise is added to the synthetic GXD (not shown), and is likely exacerbated by 3D effects \& hot-cold fuel mix so that even large positive hot spot $a_4$ on DT layered capsules may remain undetected by this hot spot x-ray self-emission technique. 
Analysis of synthetic GXD images created from the 2D Hydra runs suggest that the $a_4$ measured experimentally with the GXD is not a true representation of the hot spot $a_4$, particularly for large positive $a_4$ amplitudes. Fig. \ref{fig3}(a) depicts the relationship between the DT layered capsule ``hot spot $a_4$'' (as previously defined) and that from the 17\% contour of the synthetic GXD (the synthetic GXD $a_4$), both were extracted at x-ray bangtime (the principal value used for analysis of experimental data). $a_4$ measured from the synthetic GXD is consistently lower than that of the hot spot. The insensitivity to positive hot spot $a_4$ is caused by lateral ablator mass flows which accumulate at $\sim 45 ^{\circ}$ at the expense of ablator material near the equator and poles (see Fig. \ref{picture} (a)). The ablator material is rotationally symmetric about the horizontal axis, so the accumulated material absorbs the x-rays emitted from the polar-lobes of the hot spot (left and right), while allowing x-rays to more readily pass through the equatorial regions (top and bottom). Consequently the polar-lobes of the hot spot which are visible in the density plots of Fig. \ref{picture} as dark regions (the hot spot is the central region of low density) are almost completely invisible in the GXD plots compared to the emission through the equator. This causes the x-ray image to have a pronounced negative $P_2$ shape (oblate or ``pancaked''). As the hot spot $a_2=0\pm 1$ \textmu m ($a_2$ is the amplitude of the $P_2$ mode) for all these pure $P_4$ modelling runs, the $P_2$ inferred from the x-ray image is a ``false'' negative $P_2$ mode. This suggests that a negative $P_2$ mode measured from the self-emission x-ray image may in fact be a signature of a positive $P_4$ mode, although it does not, of course, preclude the presence of a true $P_2$ mode. This is potentially important for interpretation of x-ray images from DT implosions, which often exhibit oblate (negative) $P_2$ modes \cite{Glenzer05032010}.  

NIF experiments also use low convergence, DHe$^3$ gas filled ``symmetry capsules'' which have a surrogate CH fuel mass. Using Hydra, symmetry capsule \& DT layered capsule pairs of runs were created by applying identical x-ray drives to both capsules. When realistic experimental noise is applied to the synthetic GXD images, symmetry capsules, which have better signal to noise ratios and larger stagnation diameters, enable a far better measurement of the $P_4$ mode than equivalent DT layered capsules. Nevertheless, these calculations indicate symmetry capsules also show reduced sensitivity to $a_4$, exhibit a ``false'' $P_2$, and are quantitatively very similar to those shown in Figs. \ref{fig3} (a) \& (b) respectively.

\begin{figure}[t!]
\centering
\includegraphics[scale=.22,trim=25mm 30mm 15mm 40mm,clip]{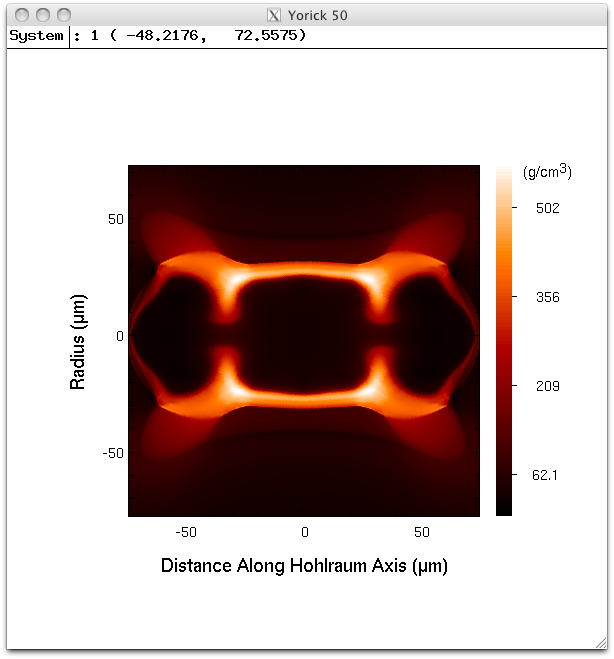}
\includegraphics[scale=.22,trim=35mm 30mm 15mm 40mm,clip]{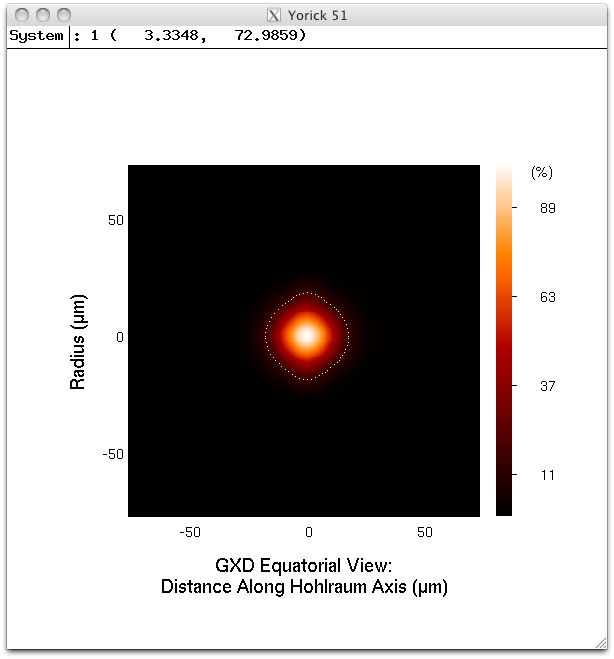}
\includegraphics[scale=.18,trim=10mm 20mm 50mm 20mm,clip,angle=0]{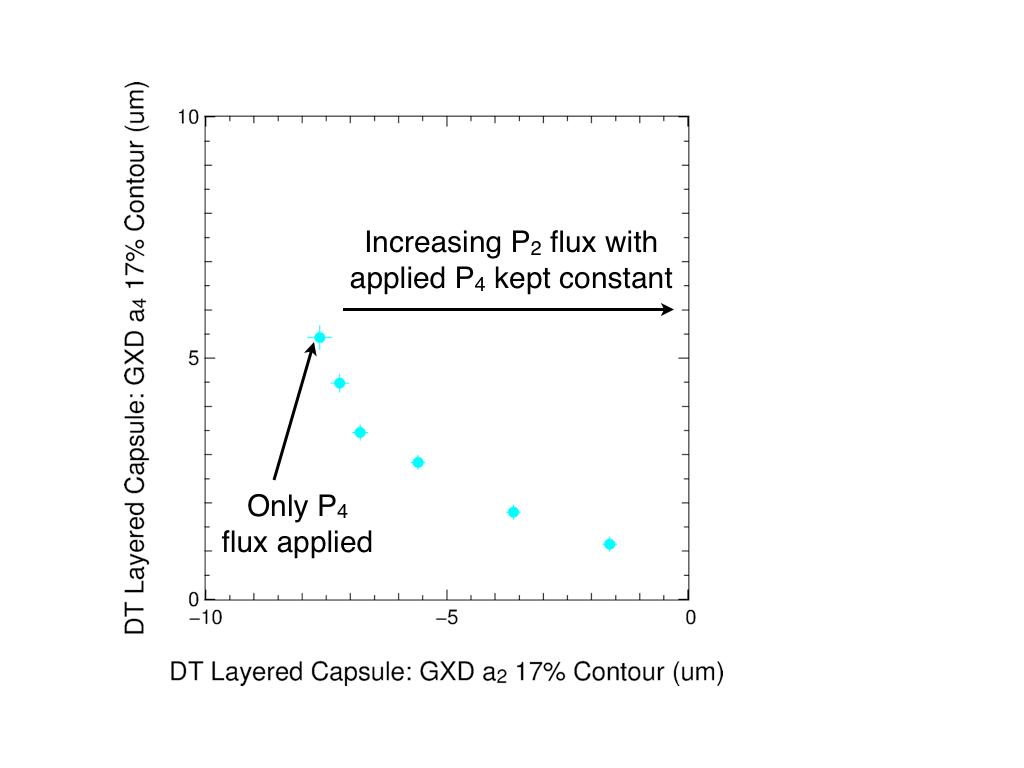}
\caption{(a) Density plot of a DT layered capsule run with both $P_2$ and $P_4$ flux modes applied. Axis of rotational symmetry is horizontal at Radius $=0$ \textmu m. (b) The equatorial synthetic GXD image of (a) at the same time, dotted line shows 17\% contour. Despite the highly non-spherical density distribution, the equatorial GXD image is almost perfectly round. (c) As the $P_2$ flux amplitude is increased in order to make the GXD image look round (reducing $a_2$) the sensitivity to $a_4$ is reduced towards zero.}
\label{fig5}
\end{figure}

DT implosions on the NIF currently have yields $\sim 3 - 10 \times$ below detailed 2D post-shot Hydra simulations \cite{Clark:2012fk} that match the measured shock timing, implosion velocity, and capsule and ice surface roughnesses. In comparison to experimentally measured or inferred values \cite{Springer:2011uq}, these simulations have similar hot spot temperatures, however the hot spot volumes are reduced while the hot spot mass is increased, causing a $2-3\times$ discrepancy in the hotspot density. $P_4$ shape perturbations offer one mechanism which may explain these experimental observations in particular bringing the yield and ion temperature relationship into better agreement. In these simulations, the DT fuel and hot spot do not mix; clear boundaries still exist (note these simulations use smooth capsules, but when nominal realistic capsule surface roughness \cite{clark:082701} was employed and modes up to 200 resolved, no significant implosion degradation occurred for the full range of $a_4$). Consequently unlike high mode `mix' \cite{lindl:339} (where the hot spot can be radiatively cooled by high $Z$ impurities), the simulated ion temperature inferred from the neutron spectrum remains unaffected at $3.9\pm 0.05$ keV for all $a_4$. The large $a_4$ does however truncate the thermonuclear burn, moving both the neutron and x-ray bangtimes earlier in time, therefore as the capsule is still converging at bangtime, the hot spot size and volume are increased. The hot spot mass decreases with positive $a_4$, bringing Hydra simulations approximately in line with experimental data, as shown in Table \ref{table1}. This compares NIF experimental data \cite{Springer:2011uq} with two Hydra implosions; one is perfectly spherical while the other has a hot spot $a_4$ of +20 \textmu m. Notable features are the significantly reduced yield, reduced pressure, reduced hot spot mass, unchanged ion temperature and increased hot spot volume. We must emphasize, however, that this should not be interpreted as conclusive evidence that a $P_4$ asymmetry is responsible for the observed reduced NIF capsule performance. Although this study has concentrated on the $P_4$ mode, it is likely that all low modes would reduce the conversion of capsule kinetic energy into hot spot pressure, and may result in similar ambiguity in the shape of the x-ray emission from the hot spot \cite{B.K.:2012uq}. To explore the issue of low mode asymmetries further, experiments using x-ray backlighters are currently being conducted on NIF to measure the implosion shape both in-flight \cite{hicks:102703} and at stagnation using Compton radiography \cite{tommasini:056309}.

%A noteworthy set of NIF shots is DT layered shot N111215 and it's four symmetry capsule repeat shots. The symmetry capsules had measured $a_4=1.7, 1.1, 2.75, 6.9$ \textmu m. Our Hydra simulations indicate these can be translated into DT layered target hotspot asymmetries of $a_4 = 4, 2, 10, 24 \pm 2$ \textmu m respectively. This layered shot had the highest yield to date ($7\times10^{14}$ neutrons). Assuming the $a_4$ amplitude of the DT implosion was in the range implied by the symmetry capsules, an equivalent spherical implosion, may have had a yield of $0.1-1.1\times10^{16}$ - entering the $\alpha$ heating regime.

%In light of the insensitivity of the hotspot self emission technique to the $P_4$ mode as described in this Letter, experiments using x-ray backlighters are currently being conducted on NIF to measure the implosion shape both in-flight \cite{hicks:102703} and at stagnation using Compton radiography \cite{tommasini:056309}. Observed $P_4$ flux asymmetries may be corrected by repointing the outer cones of beams \cite{lindl:339,jones:056315}. 

\begin{table}[t!]
\centering
\begin{tabular}{| r | c | c |c | c |c | c | c | c |}
\hline
                                     			&NIF expt.	&Hydra                      		&Hydra				\\
                                   			&range\cite{Springer:2011uq}	&($a_4 = 0$ \textmu m)		&($a_4 = 20$ \textmu m)	\\
\hline
Pressure (GBar)            			&57-81	&348	 				&115 		  			\\
Mass (\textmu g)          			&2-6.4	&8					&5.5  					\\
Density (gcm$^{-3}$)       		&22-35	&136					&69	 				\\
Volume ($\times 10^{-7}cm^3$)	&0.9-1.9		&0.6		&1.0		\\
Tion (keV)				&3.3-4.4				&3.9 					&3.9					\\
Fuel $\rho r$ (gcm$^{-2}$)	&0.77-0.98			&0.7 					&0.72 				\\
Yield (neut. $\times 10^{14}$)	&1.9-6.0	&74		&5.3		\\
\hline
\end{tabular}
\caption{A comparison of NIF DT layered capsule experimental data from 4 shots N110608-N110908 with two Hydra implosions, one spherical ($a_4=0$ \textmu m, and another with $a_4 = 20$ \textmu m. Large positive $P_4$ brings the modeled implosion observables approximately in line with the experimental data.}
\label{table1}
\end{table}

As discussed, implosions with a significant $P_4$ asymmetry can have a very apparent but ``false'' $P_2$ asymmetry in GXD images. We find that attempting to correct this ``false'' $P_2$ by increasing laser power to the hohlraum waist (capsule equator) \cite{Glenzer05032010} can lead to a round GXD image even though the correction actually produces a more distorted DT fuel ice layer. This is depicted in fig. \ref{fig5} for the case of a DT layered capsule where we applied and empirically adjusted a $P_2$ flux asymmetry, in addition to the original $P_4$, in order to make the synthetic GXD image appear round. Fig. \ref{fig5}(c) quantifies a related effect; as the applied $P_2$ flux is increased in order to reduce the ``false'' GXD $a_2$ towards zero, there is a marked additional reduction in sensitivity to $a_4$ (relative to that shown in Fig. \ref{fig3}). This suggests that attempts to tune the hohlraum to eliminate a ``false'' $P_2$ can have the unintended consequence of exacerbating overall asymmetry. Other information, such as comparison of the widths of images taken from both the polar and equatorial lines of sight \cite{benedetti:2012} need to be taken into consideration. These simulations show that when a hotspot has a positive but pure $P_4$ asymmetry the equatorial image width is larger than the width in the polar image (for negative $P_4$ this is reversed). This could be used to identify an implosion where the measured $P_2$ may be caused by a dominant $P_4$ asymmetry. However, our simulations also show us that the empirically but incorrectly tuned implosion of fig. \ref{fig5} would have a polar image width that is equal to the equatorial image width, further misleading us into thinking that we had engineered an approximately spherical implosion. A corollary of figure \ref{fig5}, is that it is possible to create imploded configurations which appear to be symmetric in the GXD but, in fact, are significantly asymmetric and have greatly reduced performance in comparison to equivalent spherical implosions because a large fraction of the imploding shell's kinetic energy remains unstagnated.

In summary, numerical simulations have been used to examine the sensitivity of implosions similar to those currently taking place on NIF to low-mode flux asymmetries. It is shown that Legendre polynomial $P_4$ flux modes induce $P_4$ shape modes at the time of capsule stagnation. The largest $P_4$ amplitudes can cause up to 50\% of the capsule kinetic energy to remain unconverted to hot spot and DT ice internal energy, in turn reducing the neutron yield by up to $15\times$. Simulated x-ray images of the hot spot self-emission show reduced sensitivity to the positive $P_4$ mode, instead the images appear to have a pronounced oblate $P_2$ shape. Attempting to correct for this apparent $P_2$ distortion can further distort the implosion while creating x-ray images which appear round \& self-consistent from both equatorial and polar directions. This also further reduces the sensitivity to the $P_4$ mode such that that no quantitative evaluation of the hot spot $a_4$ can be made. Long wavelength asymmetries may be playing a significant role in the observed yield reduction of NIF DT implosions relative to detailed post-shot 2D simulations.

\end{document}